\documentclass[]{emulateapj}
\usepackage{graphicx}
\usepackage{amssymb}
\usepackage{amsmath}

\begin{document}

\title{Toward Epoch of Reionization Measurements with Wide-Field Radio Observations}
\author{Miguel F. Morales}
\affil{MIT Center for Space Research\\37-664H, 77 Massachusetts Ave.\\Cambridge, MA 02139}
\email{mmorales@space.mit.edu}
\and
\author{Jacqueline Hewitt}
\affil{MIT Center for Space Research\\37-241, 77 Massachusetts Ave.\\Cambridge, MA 02139}
\email{jhewitt@space.mit.edu}

\begin{abstract}

This paper explores the potential for statistical epoch of reionization (EOR) measurements using wide field radio observations. New developments in low frequency radio instrumentation and signal processing allow very sensitive EOR measurements, and the analysis techniques enabled by these advances offer natural ways of separating the EOR signal from the residual foreground emission. This paper introduces the enabling technologies and proposes an analysis technique designed to make optimal use of the capabilities of next generation low frequency radio arrays.  The observations we propose can directly observe the power spectrum of the EOR using relatively short observations, and are significantly more sensitive than other techniques which have been discussed in the literature.  For example, in the absence of foreground contamination the measurements we propose would produce five 3-sigma power spectrum points in 100 hours of observation with only 4 MHz bandwidth with LOFAR for simple models of the high redshift 21cm emission. The challenge of residual foreground removal may be addressed by the symmetries in the three-dimensional (two spatial frequencies and radiofrequency) radio interferometric data.  These symmetries naturally separate the EOR signal from most classes of residual un-subtracted foreground contamination, including all foreground continuum sources and radio line emission from the Milky Way.

\end{abstract}

\section{Introduction}

The possibility of observing 21 cm emission from the ``cosmic dark ages'' was first recognized by \cite{SunyaevZeldovich} and later developed by \citet{ScottRees}, \citet{MMR97}, and \citet{Tozzi} to show the nature of the expected signal and demonstrate that future low frequency radio telescopes could measure this emission. This has inspired a significant amount of work, with important contributions by \citet{Iliev}, \citet{DiMatteoForegrounds}, \citet{Carilli21Absorption}, \citet{FurlanettoMinihalos} and others. Just recently, \citet{RennanNoStep} showed that the predicted global step in reshifted 21 cm emission was an artifact of the small box size used in the simulation procedures.  Combined with the high Thomson scattering cross section observed by WMAP \citep{WMAPTE} and other arguments, these results suggest that reionization is much more complicated than originally conceived \citep{HuiThermalMemory}.

These theoretical developments coincide with the development of several advanced low frequency radio arrays
which will come online by the end of the decade.  Recent advances in digital electronics are enabling the construction of radio arrays which directly sample the incoming wavefront with thousands of simple antennas and perform all signal processing in the digital domain.  The digital processing simplifies the mechanical systems, allowing very large collecting areas at modest cost and enabling entirely new observing modes.

One of the enabling digital processing technologies is the Wide Field Correlator (WFC) developed at MIT and the Haystack Observatory \citep{WFC}.  The digitized signal from up to several thousand individual antennas can be fully correlated by the WFC, producing more than 1 billion individual visibility measurements. These visibilities allow the full field of view of the antennas to be imaged, and greatly increases the sensitivity of epoch of reionization (EOR) observations by allowing both parallel observation of a large number of sky locations and by enabling new statistical analysis techniques (see Section \ref{AnalTechSection}).

In this paper we show how wide field radio observations combined with statistical analysis techniques similar to those used in CMB experiments may be applied to studies of the EOR. An analysis using statistical CMB techniques independently developed by \citet{ZaldariaggaPow1} has been recently published, where they show the usefulness of the EOR power spectrum and that foreground contamination can be easily removed. Our approach is similar but emphasizes the importance of the full three dimensional statistics inherent in the EOR signal. By making full use of all the available information, our technique is both more sensitive to the EOR signal and provides additional handles for removing the foreground contamination. This paper complements the \citet{ZaldariaggaPow1} paper and details the experimental and analysis advances which will enable future radio arrays to make robust EOR measurements using reasonable integration times.

This discourse is divided into three major themes.  In Section \ref{WFCSection} the basic design and wide field capabilities of future low frequency radio arrays are presented, followed by a description of the analysis technique enabled by the wide field of view in section \ref{AnalTechSection}.  Sections \ref{PowerSpectrumSection} and \ref{SimpleCaseSection} then present sensitivity calculations for wide field radio EOR observations and the constraints they can provide.  Additional calculations and issues important to the EOR field are detailed in the appendices.

\section{Wide Field Radio Observations}
\label{WFCSection}

In contemporary and planned radio telescopes, the digitization happens as early as possible in the signal collection process and signal processing is performed in high speed digital electronics.  This shift to early digitization dramatically affects the ``look'' of the array, with thousands of small elements replacing a few large parabolic antennas.  This basic change in the design philosophy enables wide field observations but also places much greater demands on the backend digital processing. This section introduces the basic antenna design and the massively interconnected correlators that can scale to thousands of antennas. While none of the proposed arrays have been constructed, they all share the same basic design elements which are used as the basis for the following discussion.

\begin{figure*}
\begin{center}
\plotone{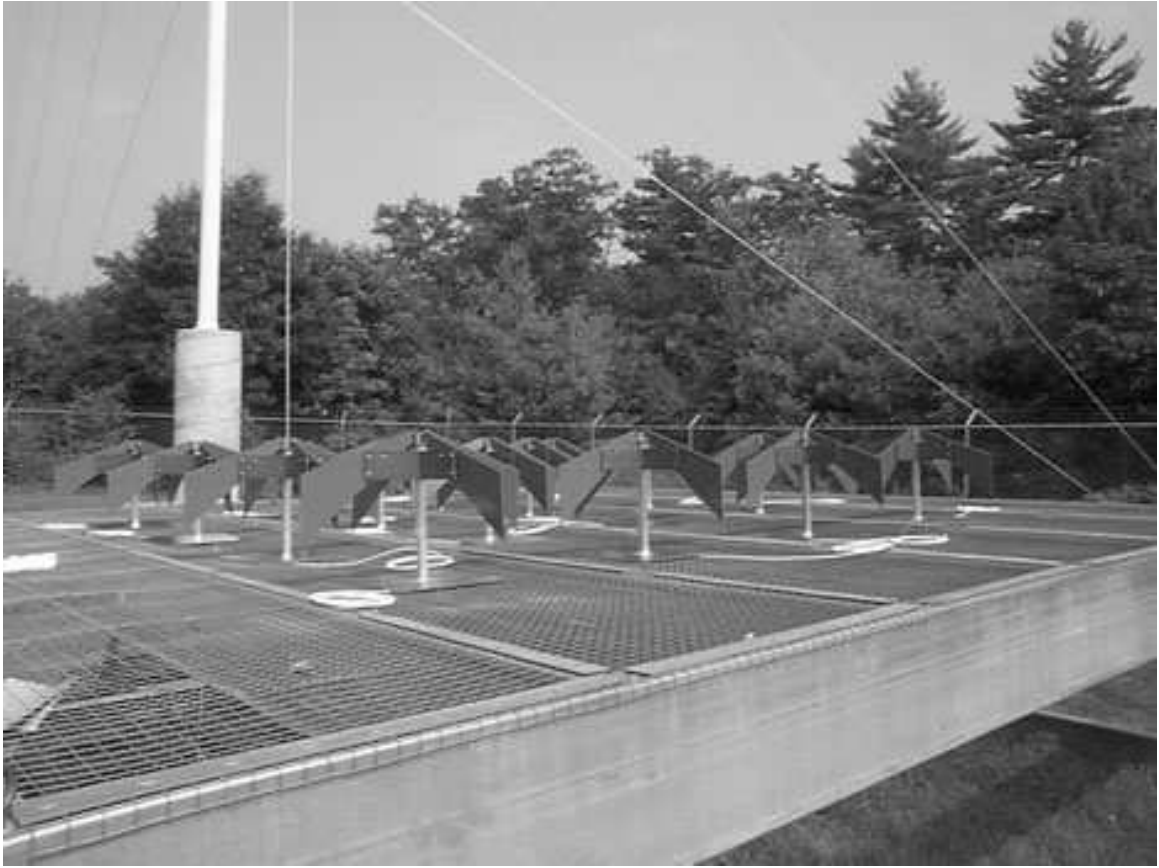}
\caption{This photograph shows a prototype for one sixteen element antenna for the 80 -- 240 MHz band. }
\label{HBantenna}
\end{center}
\end{figure*}

The individual receptor elements consist of simple dipoles.  Dipoles may then be grouped with an analog beamformer into antenna ``tiles.'' The more dipoles per tile, the larger the collecting area and smaller the field of view.  Depending on the the science goals, the number of dipoles per antenna is then chosen to optimize science return and cost.  The 80 -- 240 MHz prototype tile developed by the MIT Haystack Observatory consist of 16 crossed dipoles, maximizing the collecting area at frequencies critical for EOR studies at the expense of limiting the field of view to $20^{\circ}$ (see Figure \ref{HBantenna}) .  After each antenna the signal is immediately digitized and all subsequent processing is done in the digital domain.

Wide field observations which image the full antenna field of view are enabled by the Wide Field Correlator technology \citep{WFCpaper}.  Since the number of visibilities grows as the square of the number of antennas, arrays with thousands of antennas require millions of baselines, necessitating a massively interconnected correlator technology.  The WFC technology enables large arrays to be fully correlated at low cost by developing scalable techniques for handling millions of pairwise connections with a minimum of point-to-point connections.  The WFC technology also addresses a number of additional technological issues to allow efficient use of FPGA chips including time multiplexing to match bandwidth with chip speed, and data buffering and re-ordering to minimize on chip memory requirements.  This architecture is very scalable, and the performance of future versions of the WFC should increase with Moore's law providing even greater performance at very reasonable cost.  

The first generation wide field correlator (WFC1) performs the pairwise correlations for 3072 antennas at 4 MHz bandwidth,  producing 1.2 billion 16 kHz visibilities (see \citet{WFC} for a full description).  The capacity of an FX correlator design can be approximated by the trillions of complex multiply and accumulates per second (Tcmacs) performed, given by the number of baselines (pairwise combinations of antennas) times the bandwidth. The WFC1 has a performance of 38 Tcmacs ( $(3072^{2}/2){\rm\ baselines}\times 4 {\rm\ MHz} \times 2 {\rm\ pol}= 38 {\rm\ Tcmacs}$), and is expected use $\sim 2700$ FPGA chips on 42 boards with a hardware cost of under 1 million dollars US (2006 price).  For massively interconnected correlators such as the WFC1, the number of antennas can be traded for bandwidth with little additional cost as long as the number of Tcmacs remains constant.


The baseline observing program for wide field EOR observations involves a few hundred hours of observations in a cold portion of the sky.  While all of the foreground sources will be subtracted from the data, residual foreground contamination can be reduced by carefully choosing the observation direction.  The observation location will be away from the galactic plane to reduce the strength of the galactic emission, and will be chosen to contain a minimum of bright sources in an area where the galactic emission is particularly smooth.  The observing times are selected when the chosen location is at high elevation angle and the ionospheric conditions are particularly favorable.  In practice this will mean observing for a couple of hours on many nearly consecutive nights when ionospheric conditions are at their best.  With six months of observing, a very clean data set can be obtained with a deep integration of a few hundred hours.

The combination of very large effective area and wide field of view makes next generation low frequency observatories ideally suited to EOR measurements. While the wide field of view can be thought of in the traditional sense as allowing $\sim 1$ hundred thousand simultaneous pointings, in the context of EOR observations it is more productive to think in terms of the visibility measurements.  As will be presented in the next section, the EOR information is most easily extracted in the Fourier representation, and one can think of the visibilities produced by the WFC1 as 1.2 billion independent measurements.  This large number of independent measurements allows wide field radio observations to directly sample the statistical signature of the EOR.

\section{Statistical Analysis Technique}
\label{AnalTechSection}

In this section we introduce our statistical analysis of EOR measurements in three parts.  In the first part we introduce the formalism for calculating the three dimensional statistics of the EOR signal.  This is followed by a discussion of the inherent symmetries of the statistics and how this can be used to separate residual foreground contamination from the EOR signal in Section \ref{SymmetrySubSection}, and a discussion of the non-Gaussian properties of the statistics in Section \ref{NonGaussianSubSection}.  The formalism presented in Section \ref{VisStatIntro} parallels calculations for the CMB by \citet{White} extended to three dimensions, and is similar to calculations by \citet{ZaldariaggaPow1} and \citet{BharadwajVis} but uses Fourier formalism to clearly illustrate the three dimensional nature of the problem and to highlight the symmetries and statistics which are explored in Sections \ref{SymmetrySubSection} and \ref{NonGaussianSubSection}.  The idea of using power spectra to measure the EOR signal has been around for some time, but the power of the symmetries inherent in the three dimensional data set and the non-Gaussian statistical properties of the signal have not been previously appreciated.

\subsection{Introduction to the Visibility Statistics of the EOR}
\label{VisStatIntro}
In wide field radio observations, the observed frequency of the redshifted 21 cm line maps to the line of sight distance of the source. While peculiar velocities mean this mapping is not exact---a small effect to be studied in future work---an approximate image cube can be formed by replacing the observed frequency with the equivalent line of sight distance.  The resulting image cube is a full three dimensional tomograph of neutral hydrogen emission in the high redshift universe.

\begin{figure*}
\begin{center}
\plotone{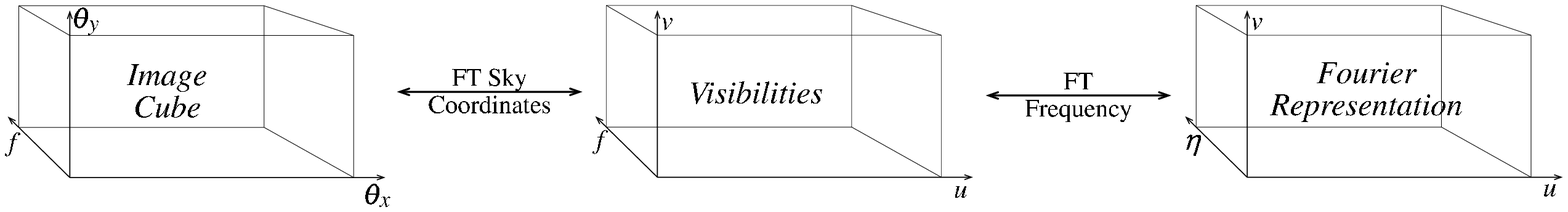}
\caption{This diagram shows the Fourier transform relationships between the image cube, the measured visibilities, and the Fourier representation.  For an interferometer, the fundamental observable is the visibility-frequency cube, which can then be transformed into either an image cube to show source locations or the Fourier representation to analyze the spatial structure of the signal.}
\label{cubeTransform}
\end{center}
\end{figure*}

The emission strength from neutral hydrogen during the epoch of reionization depends on the local hydrogen density, the fractional ionization, and the spin temperature of the gas, and gives us a differential specific intensity $\triangle I$ at the observed location with respect to the cosmic microwave background radiation (see Appendix \ref{AppendixBD} for a full derivation of $\triangle I$).  Because the 21 cm line is narrow and weak and the gas is optically thin, there is no significant absorption of background emission by neutral hydrogen nearer the observer.  This allows us to directly form an image cube which represents the three dimensional emission of the neutral hydrogen.  Furthermore, a three dimensional Fourier transform can be applied to the image cube to determine the spatial structure of the EOR signal.  This gives us four representations of the signal:
\begin{equation}
\label{Fsquare}
\begin{array}{ccc}
\triangle I(\vec{r}) &{\Leftarrow\!\Rightarrow} & \triangle \tilde{I}(\vec{k}) \\
\\
\Bigg\downarrow& &\Bigg\downarrow \\
\\
\triangle I(\theta_{x},\theta_{y},\triangle\! f) & \Leftarrow\!\Rightarrow &\triangle \tilde{I}(u,v,\eta).
\end{array}
\end{equation}
The horizontal pairs in Equation \ref{Fsquare} are related by Fourier transforms and the vertical pairs are related by a change of variables from the observational quantities sky position $\theta_{x},\theta_{y}$ and frequency $\triangle\! f$ (and their inverses $u,v,\eta$) to the comoving distance $\vec{r}$ (and inverse $\vec{k}$). The upper left hand term is the brightness distribution of the 21 cm emission at the source while the lower left hand term is the observed image cube.  Similarly the upper right hand term gives the spatial structure of the EOR signal and the lower right hand term is the spatial structure of the three dimensional image.  The mapping between variables is given by
\begin{eqnarray}
&\theta_{x} =  \frac{r_{x}}{D_{M}(z)}, &  u =\frac{k_{x}{D_{M}(z)}}{2\pi}, \label{thetar}\\
&\theta_{y} =  \frac{r_{y}}{D_{M}(z)}, & v = \frac{k_{y}{D_{M}(z)}}{2\pi},\\
&\triangle\! f \approx  \frac{H_{0}f_{21}E(z)}{c\,(1+z)^{2}}\, \triangle r_{z}, & \eta \approx \label{fapprox}\frac{c\,(1+z)^{2}}{2\pi H_{0}f_{21}E(z)} k_{z},
\end{eqnarray}
where $f_{21}$ is the frequency of the 21 cm line, $D_{M}$ is the transverse comoving distance \citep{Hogg}, $E(z) \equiv \sqrt{\Omega_{\rm M}(1+z)^{3} + \Omega_{\rm k}(1+z)^{2} + \Omega_{\Lambda}}$, and $z$ is the reference redshift at the middle of the observed data cube used for determining $\triangle f$ and $\triangle r_{z}$.\footnote{\label{FTnote}The top line of Equation \ref{Fsquare} is a familiar relationship from cosmology and the bottom line is a standard of radio astronomy.  Unfortunately the two different communities use different conventions for the Fourier transform.  In this paper we will use the transformation pair
\begin{equation}
\label{CFT1}
\triangle I(\vec{r}) = \int  \triangle \tilde{I}(\vec{k}) e^{i\vec{k}\cdot\vec{r}}d^{3}\vec{k},
\end{equation}
\begin{equation}
\label{CFT2}
\triangle \tilde{I}(\vec{k}) = \frac{1}{(2\pi)^{3}} \int  \triangle I(\vec{r}) e^{-i\vec{k}\cdot\vec{r}}d^{3}\vec{r}, 
\end{equation}
for transforming functions of position $\vec{r}$ and $\vec{k}$ and
\begin{equation}
\triangle I(\theta_{x}, \theta_{y},\triangle\! f) = \iiint  \triangle \tilde{I}(u,v,\eta) e^{i 2 \pi (u\theta_{x}\hat{\i}+v\theta_{y}\hat{\j}+\eta\triangle f\hat{k})}du\, dv\, d\eta, 
\end{equation}
\begin{equation}
\triangle \tilde{I}(u,v,\eta) = \iiint  \triangle I(\theta_{x}, \theta_{y},\triangle\! f) e^{-i 2 \pi (u\theta_{x}\hat{\i}+v\theta_{y}\hat{\j}+\eta\triangle\! f\hat{k})}d\theta_{x}\, d\theta_{y}\, d\triangle\! f, 
\end{equation}
for functions of the observation variables.  While using two different definitions of the Fourier transform is awkward, it allows us to use the standard notation of both cosmology and radio astronomy and avoids the inevitable confusion that would result from non-standard notation. It is this difference in transforms which explains the relative $2\pi$ factors in Equations \ref{thetar}--\ref{fapprox}. Standard units are comoving Mpc for $r$, rad per comoving Mpc for $k$, radians for $\theta$, Hz for $\triangle f$, wavelengths (or equivalently rad$^{-1}$) for $u$ and $v$, and Hz$^{-1}$ for $\eta$.  If $r$ and $k$ are in proper coordinates instead of comoving coordinates, Equations \ref{thetar}--\ref{fapprox} need to be divided by $(1+z)$. Note that while $\eta$ physically has the units of seconds, it is really acting as a stand-in for the spatial structure and is best thought of as an inverse distance.}$^{\rm ,}$\footnote{An exact derivation maps the observed frequency $f$ to 
\begin{eqnarray}
\label{ }
\theta_{f}= \frac{c}{H_{0}}\int_{f_{\rm ref}}^{f}\frac{f'^{2}df'}{f_{21}D_{M}(f')E(f')}.
\end{eqnarray}
$\theta_{f}$ is then an exactly linear map to the line of sight distance and has the same scale as $\theta_{x}$ and $\theta_{y}$.  This alternate development leads to a kernel which must be added to the Fourier transform relationship developed in Equation \ref{visftrans} and Figure \ref{cubeTransform}.  For almost all reasonable parameters the approximate relationship in Equation \ref{fapprox} is sufficient.} 


For an interferometric array the fundamental observable is the set of cross-correlations between the antenna elements $V(u,v,\triangle\! f)$, called visibilities in radio astronomy, and is related to the image cube and the Fourier representation via Fourier transforms.  The Fourier representation of the image cube $\triangle \tilde{I}(u,v,\eta)$ is related to the measured visibilities via the one dimensional Fourier transform along the frequency axis
\begin{equation}
\label{visftrans}
\triangle \tilde{I}(u,v,\eta) = \int V(u,v,\triangle\! f)\, e^{-2\pi i \,\triangle\! f \eta}\, d\triangle\!f,
\end{equation} 
and the image cube is obtained by performing a two dimensional Fourier transform in the sky coordinates
\begin{equation}
\label{ }
\triangle I(\theta_{x}, \theta_{y},\triangle\! f) = \iint  V(u,v,\triangle\! f) e^{i 2 \pi (u\theta_{x}\hat{\i}+v\theta_{y}\hat{\j})}du\, dv.
\end{equation}
These relationships are shown graphically in Figure \ref{cubeTransform}. 

In order to calculate the actual measured values of $\triangle \tilde{I}$, $\triangle I$, and $V$, we need to understand the mapping from the true intensity distribution $\triangle I_{T}$ to the measured quantities. In particular, the finite field of view and bandwidth of the measurement truncates the Fourier transform from the true brightness distribution to the measured visibilities.

In what follows we neglect the curvature of the sky.  For the purposes of examining the statistics of radio emission on the celestial sphere, this assumptions is valid when $2 sin(\theta/2) \approx \theta$ \citep{BondEfstathiou}. For our proposed observations the largest value of $\theta$ is $10^\circ$, and we can adopt the two dimensional Fourier relationships rather than the full statistical treatment of spherical harmonics. The condensed arrays are expected to be coplanar to a fraction of a wavelength at EOR frequencies, so the ``w-term'' from relative antenna elevations may be neglected.  Additionally, the visibilities from the Wide Field Correlator will be collected every 1/2 second and the baseline adjusted to allow tracking of a single patch of the radio sky.

The true distribution of the EOR specific intensity is given by
\begin{equation}
\label{trueVis}
\triangle I_{T}=F_{\rm HI} \equiv C \,\frac{\rho_{\rm HI}}{\left \langle \rho_{\rm H}\right \rangle},
\end{equation}
where $F_{\rm HI}$ gives the fluctuations in the neutral hydrogen emission. This can be separated into a spin temperature component $C$ and the density terms, where $\rho_{\rm H}$  and $\rho_{\rm HI}$ are the density of hydrogen and neutral hydrogen respectively.  For a baryon density $\Omega_{\rm B}h^{2} = 0.02$ and a helium fraction $Y = 0.24$, $C$ is given by 
\begin{equation}
\label{ }
C_{\rm Jy} \equiv (2.9\ {\rm mK})\,h^{-1}\frac{\left(I_{\rm HI}-I_{\rm CMB} \right)}{E(z)\, T_{s}}
\end{equation}
when measured in Jy str$^{-1}$, and
\begin{equation}
\label{ }
C_{\rm K} \equiv (2.9\ {\rm mK})\,h^{-1}\frac{(1+z)^{2}} {E(z)}\frac{\left(T_{\rm s}-T_{\rm CMB} \right)}{T_{s}}
\end{equation}
in brightness temperature units.  $C$ captures the spin temperature dependence of the neutral hydrogen emission and $\rho_{\rm HI}/\left \langle \rho_{\rm H}\right \rangle$ includes the effects of density fluctuations and ionization fraction, and in general both terms are functions of position.\footnote{We have followed EOR convention and defined $\triangle I_{T}$ with respect to the CMB radiation and not the CMB radiation plus the average contribution of the neutral hydrogen emission.  This choice of zero point differs from the convention used in derivations of CMB fluctuations, and can add a $\delta$-function at zero wavenumber to the power spectrum.  This offset is very hard to measure in practice, and its contribution is ignored in the remainder of this paper.  Mathematically the zero point contribution can be removed by subtracting $C\left \langle \rho_{\rm HI}\right \rangle/\left \langle \rho_{\rm H}\right \rangle$ from Equation \ref{trueVis} to redefine $\triangle I_{T}$.} A full derivation of $C$ along with definition of all terms is given in Appendix \ref{AppendixBD}.

The first step in determining the measured quantity is to define the window function $W(\theta_{x},\theta_{y},f)$ given by the instrumental field of view and bandwidth.  This can be Fourier transformed to obtain $W(u,v,\eta)$.  Since the measured brightness $\triangle I(\theta_{x},\theta_{y},f)$ is given by multiplying the true brightness by the spatial window function, the measured Fourier transform of the intensity is given by convolving the Fourier transform of the true intensity $\triangle \tilde{I}_{T}(u,v,\eta)$ with $W(u,v,\eta)$ 
\begin{multline}
\label{trueIconv}
\triangle \tilde{I}(u_{1},v_{1},\eta_{1}) = \\ \int \triangle \tilde{I}_{T}(u,v,\eta)W(u_{1}-u,v_{1}-v,\eta_{1}-\eta) \,du\,dv\,d\eta.
\end{multline}
Using Equation \ref{trueVis} this can be rewritten as
\begin{multline}
\label{measvis1}
\triangle \tilde{I}(u_{1},v_{1},\eta_{1}) = \\ \int F_{\rm HI}(u,v,\eta)W(u_{1}-u,v_{1}-v,\eta_{1}-\eta)\,du\,dv\,d\eta
\end{multline}
for a specific realization of $F_{\rm HI}$. Because model predictions of $F_{\rm HI}$ are statistical it is convenient to consider the average value.  For the next several steps we will first sketch the derivation of the statistical relationship between the intensity and the neutral hydrogen fluctuations in $\vec{n}$ space, where $\vec{n}$ is a wave number vector in undefined units, but is used to stand in for $\vec{k}$ or $u\hat{\imath} + v \hat{\jmath} + \eta\hat{k}$.  After working through the derivation in the abstract $\vec{n}$ space we will redo the analysis in physical units. Rewriting Equation \ref{measvis1} in the $\vec{n}$ space we obtain
\begin{equation}
\label{visWink}
\triangle \tilde{I}(\vec{n}') = \int \! F_{\rm HI}(\vec{n})\, W(\vec{n}'-\vec{n})\,d^{3}\vec{n}
\end{equation}
We can use this to calculate the expectation value of two visibilities measured at two different locations $\vec{n}'$ and $\vec{n}''$
\begin{multline}
\label{}
\left\langle \triangle \tilde{I}(\vec{n}')\,\triangle \tilde{I}^{\ast}\!(\vec{n}'')\right\rangle = \\ \Bigg\langle \int\!\! F_{\rm HI}(\vec{n})\,W(\vec{n}'-\vec{n})\,d^{3}\vec{n}\ \\ \cdot \int \!\! F_{\rm HI}^{\ast}\!(\vec{n}''')\,W^{\ast}\*(\vec{n}''-\vec{n}''')\,d^{3}\vec{n}'''\Bigg\rangle.
\end{multline}
The model expectation for the fluctuations is spatially homogeneous (no preferred point in space), so the expectation for the Fourier transformed value $F_{\rm HI}(\vec{n})$ has random phase.  The statistical average $\left \langle F_{\rm HI}(\vec{n})\,F_{\rm HI}^{\ast}\!(\vec{n}''') \right \rangle =  \left \langle |F_{\rm HI}(\vec{n}''')|^{2} \right \rangle \delta(\vec{n}-\vec{n}''')\,d^{3}\vec{n}'''$ due to the random phase, and this is not changed by multiplying by a deterministic function like $W$.  So the expectation value gives us:
\begin{equation}
\label{viscorl}
\left\langle \triangle \tilde{I}(\vec{n}')\,\triangle \tilde{I}^{\ast}\!(\vec{n}'')\right\rangle = \int \! P_{\rm HI}(\vec{n})\,W(\vec{n}'-\vec{n})\,W^{\ast}\!(\vec{n}''-\vec{n})\,d^{3}\vec{n},
\end{equation}
where $P_{\rm HI}(\vec{n})\equiv \left \langle |F_{\rm HI}(\vec{n}''')|^{2} \right \rangle \delta(\vec{n}-\vec{n}''')$ and is the power spectrum of neutral hydrogen emission.\footnote{Note that sometimes the power spectrum is defined as $P_{\rm HI}(\vec{n})\equiv (2\pi)^{-3}\left \langle |F_{\rm HI}(\vec{n}''')|^{2} \right \rangle \delta(\vec{n}-\vec{n}''')$ in the CMB literature due to a different definition of the Fourier transform relation (see Footnote \ref{FTnote}, Equations \ref{CFT1} and \ref{CFT2}), leading to a difference of $(2\pi)^{-3}$ in the results.}  This is the result we were looking for which relates the statistical properties of the intensity to the power spectrum of emission fluctuations, and is equivalent to the standard CMB development extended to three dimensions.  For cases of constant ionization fraction and spin temperature, the analogy is even closer and $P_{\rm HI}$ becomes simply the power spectrum of the matter density fluctuations (see Section \ref{SimpleCaseSection}). This equation also shows a number of interesting features including the correlation length of the intensities.  The easiest way to use this in an analysis is to compare the intensity of locations with themselves to obtain the power spectrum of the intensity fluctuations
\begin{equation}
\label{ }
\left\langle \left\lvert \triangle \tilde{I}(\vec{n}')\right\rvert^{2} \right\rangle =  \int \! P_{\rm HI}(\vec{n}) \left\lvert W(\vec{n}'-\vec{n}) \right\rvert ^{2}d^{3}\vec{n}.
\end{equation}

We can now repeat these steps using the observation units $u,v,\eta$. Equation \ref{viscorl} gives the expectation for the intensity at two locations:
\begin{multline}
\label{bigkahuna}
\left\langle \triangle \tilde{I}(u_{1},v_{1},\eta_{1}) \,\triangle \tilde{I}^{\ast}\! (u_{2},v_{2},\eta_{2})\right\rangle =\\ \int \! P_{\rm HI}(u,v,\eta)\,W(u_{1}-u,v_{1}-v,\eta_{1}-\eta) \times \\  W^{\ast}\!(u_{2}-u,v_{2}-v,\eta_{2}-\eta)\,du\,dv\,d\eta.
\end{multline}
For a single location this simplifies to
\begin{multline}
\label{pkconclusion}
\left\langle \big\lvert \triangle \tilde{I}(u_{1},v_{1},\eta_{1})\big \rvert^{2} \right\rangle \\ = \int P_{\rm HI}(u,v,\eta) \, \left\vert W(u_{1}-u,v_{1}-v,\eta_{1}-\eta)\right\vert^{2}du\,dv\,d\eta.
\end{multline}
For the specific case of observing a single small area of the sky,  Parseval's relation can be used to convert the convolution in Equation \ref{pkconclusion} to the integrals over the Fourier space found in Equations 12 and 13 of \citet{Tozzi}, as corrected by \citet{Iliev}.

The statistical distribution calculated in Equation \ref{pkconclusion} can be directly sampled using the 1.2 billion visibilities produced by the first generation Wide Field Correlator.  Because of the correlation length introduced by the window function, not all of the visibilities measure independent values of $\triangle \tilde{I}$ (there are $\sim 20$ million independent values of $\triangle \tilde{I}$).  However, because the visibilities are independent measurements their noise is uncorrelated,\footnote{There is some evidence of correlated power on short baselines from the VLA.  From a theoretical point of view this must be either an instrumental effect or a foreground signal such as the clustered faint galaxy population. In either case this is an effect which must be dealt with for EOR measurements (see Section \ref{SymmetrySubSection}).} and the uncertainty on values of $\triangle \tilde{I}$ which are measured by multiple visibilities is correspondingly reduced.  The large number of well constrained $\triangle \tilde{I}$ measurements enabled by wide field radio observations allow the statistical distribution in Equation \ref{pkconclusion} to be directly sampled. In the following two subsections we explore how sampling the statistical distribution can help separate the EOR signal from the residual foreground and provide additional constraints on the process of reionization.

\subsection{Symmetries and Residual Foreground Subtraction}
\label{SymmetrySubSection}

The statistics of the model predictions for the EOR signal fall under the class of spatially {\em homogeneous} and {\em isotropic} random fields, since there is no preferred position or direction in space.  The underlying isotropic symmetry of the random field can be used to set strong constraints on the expected signal properties. In the limit that the observed data are from the same epoch, the isotropy of the model implies that the Fourier transform representation of the neutral hydrogen emission will be spherically symmetric since the expectation is the same in all directions. The spherical symmetry of the EOR signal is an extraordinarily useful characteristic, and all the statistical characteristics of the signal will share this symmetry (including all moments of the distributions, not just the power spectrum distribution developed in the previous section, see Section \ref{NonGaussianSubSection}).  The spherical symmetry can be used to differentiate the residual foreground signal from the EOR and to directly measure the statistical properties of the observed signal.  

Due to the spherical symmetry, all of the observations within a spherical shell of Fourier space are drawn from the same statistical ensemble.  We can thus combine all of the measurements within a spherical shell in Fourier space to determine the underlying statistical distribution. The spherical symmetry is only approximate because the universe does evolve.  However, the 4 MHz bandwidth of the baseline observations correspond to less than $4\%$ change in redshift, over which the approximation of no evolution holds quite well \citep{ZaldariaggaPow1}. Techniques which take into account the expected deviations from spherical symmetry may be employed in the future.

Differentiating the very weak EOR signal from the foreground emission of the Milky Way and extragalactic sources is one of the primary difficulties with the EOR measurement.  Planned EOR observations all involve selecting relatively cold and featureless portions of the Milky Way at high galactic latitude and carefully subtracting out the contributions of all known sources.  However, the EOR signal is so weak that residual subtraction errors and very faint foreground sources could still mask the desired signal \citep{DiMatteoForegrounds}.  The question becomes how to separate the EOR signal from the residual foreground contamination.

\begin{figure}
\begin{center}
\plotone{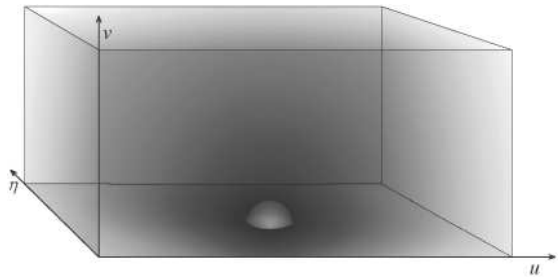}
\caption{Cartoon of the $\left \langle I(u,v,\eta)\right \rangle$ distribution of the EOR signal. Notice the spherical symmetry of the EOR signal as compared to the strong separable-axial symmetry of the residual foreground sources as shown in Figures \ref{galfore} and \ref{diffore}.  This symmetry difference can be used as a handle for separating the EOR signal from the residual foreground contamination.}
\label{eor}
\end{center}
\end{figure}

Fundamentally, the spherical symmetry of the EOR signal is based on the combination of narrow line emission and redshift which allows one to form an isotropic image cube, where frequency maps to the line of sight distance (see Figure \ref{eor}). This mechanism is not present for most of the foreground contaminants, which either have smooth continuum spectra or are local line sources.  As discussed in the following paragraphs, both of these foreground types have a \emph{separable-axial} symmetry where the specific intensity $\tilde{I}_{F}(u,v,\eta)$ for the foreground has the form $A(\eta)B(\sqrt{u^{2}+v^{2}})$ (functions $A$ and $B$ are both complex).  We can use the difference in symmetry between the spherical EOR signal and the separable-axial residual foregrounds to distinguish the EOR signal from the foreground contaminants.

\begin{figure}
\begin{center}
\plotone{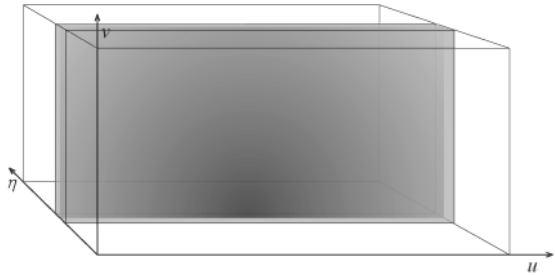}
\caption{Cartoon of the $\left \langle I(u,v,\eta)\right \rangle$ distribution of the residual faint galaxy foreground. While there is significant power in the spatial directions $u$ and $v$ \citep{DiMatteoForegrounds}, the spectral smoothness of the faint galaxy residual foreground leads to a strong separable-axial symmetry with the contribution concentrated at small $\eta$ values.}
\label{galfore}
\end{center}
\end{figure}

The most problematic smooth spectrum foreground consists of faint emission from galaxies at cosmological distances \citep{DiMatteoForegrounds, ZaldariaggaPow1}. The spectra of these sources is smooth in frequency (no sharp lines) and the spectral power law and curvature characteristics are independent of look direction.  In the image cube, these sources map to all frequencies at a given spatial location, producing pencil lines of emission along the frequency axis (the contribution need not be uniform along the pencil beam). Spatial clustering of these foreground sources can lead to features in the $u$ and $v$ Fourier coordinates, but not in the $\eta$ coordinate since the emission does not map to a single frequency.  This creates the separable-axial symmetry in the Fourier space for the dominant emission from faint unsubtracted point sources (see Figure \ref{galfore}).

\begin{figure}
\begin{center}
\plotone{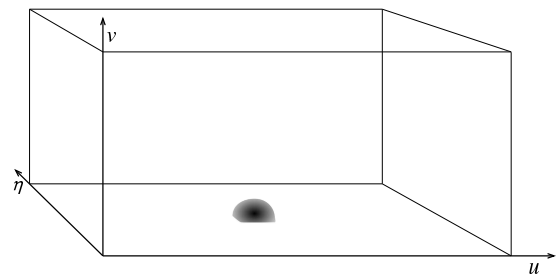}
\caption{Cartoon of the $\left \langle I(u,v,\eta)\right \rangle$ distribution of the residual Milk Way synchrotron foreground. The contribution tends to low values because the residual Milky Way synchrotron foreground is smooth both spectrally and spatially, and has a separable-axial symmetry due to the spectral characteristics.}
\label{diffore}
\end{center}
\end{figure}

The strongest foreground signal is the diffuse synchrotron radiation from the Milky Way, though this foreground is probably one of the easiest to remove \citep{ DiMatteoForegrounds}. Because this emission is fairly smooth in both angle and frequency, the residual errors from this emission is concentrated at small wavenumbers in the Fourier plane, as shown in Figure \ref{diffore}.  There may well be weak small scale structure due to clumps in the galactic emission and free-free absorption, and the correlation properties of this weak small scale structure is unknown. However, since the emission is broadband the resulting contamination has the same separable-axial symmetry in the Fourier representation as the smooth spectrum point sources.

A more difficult local foreground is produced by radio recombination lines from our own galaxy.  Line emission will also be included in the foreground model and subtracted, but significant errors in predicting and removing the recombination lines is expected.  However, in the image cube the residual recombination lines will appear as planes at set frequencies (the magnitude and sign of a residual line may vary with position, but the frequency is set).  This again leads to the separable-axial symmetry in the Fourier cube.  

Not all of the foregrounds have a separable-axial symmetry---any line source at cosmological distances will partially mimic the spherical symmetry of the EOR signal.  Fortunately, no strong lines at frequencies near or below the 21 cm line are expected in the foreground signal.  Sources such as radio recombination lines in high redshift galaxies will be a confusing source, but this emission is expected to be very weak and will have a slightly elliptical symmetry due to differences in the frequency to redshift scaling.  

Additionally, the symmetry argument does not remove the issue of foregrounds. Large residual foreground subtraction errors can swamp the EOR signal regardless of symmetry; the symmetry simply provides an additional handle for separating the foreground from the signal.  The accuracy to which a given foreground needs to be subtracted depends on both its strength and the exact shape of the Fourier signal from that source.  For example, unsubtracted discreet foreground galaxies are relatively bright, but because they are spectrally smooth their contribution is largely confined to small $\eta$ values. Radio recombination lines on the other hand are relatively weak, but due to the sharp spectral features may contain much more power at high $\eta$ values, though this depends critically on the characteristics of the radio recombination line subtraction process.

The actual contribution of a residual foreground to the Fourier representation can be directly calculated given both a model for the foreground signal and the foreground subtraction process. Returning to the visibility cube $V(u,v,\triangle f)$, let us consider the errors made along the frequency dimension for fixed $u$ and $v$. For smooth spectrum sources the two largest errors are going to be constant (complex) offsets at all frequencies due to residual source subtraction errors, and residual spectral slopes due to errors determining the spectra of the foreground sources.  This suggests a Taylor expansion of the residual errors along the frequency axis, with source subtraction errors providing the constant term, spectral index the first order term, and spectral curvature the higher order terms. The magnitude of the contamination is given by the probability of making a given error $P_{n}(\epsilon)$, where $\epsilon$ is the complex error and $n$ is the order of the Taylor expansion in the frequency dimension (in general $P_{n}(\epsilon)$ can be a function of $u$ and $v$). For the visibilities the error is given by sampling from the probability distribution $P_{n}(\epsilon)$ at each order
\begin{equation}
\label{visError}
V'(u,v,\triangle f) = V(u,v,\triangle f) +  \sum_{n=0}^{\infty} P_{n}(\epsilon) \, \epsilon\,\triangle f^{n},
\end{equation}
where $V$ is the visibility from the EOR signal and $V'$ includes the residual foreground subtraction errors. Equation \ref{visError} can be Fourier transformed along the frequency axis to give the contamination observed in the Fourier representation
%
\begin{equation}
\label{FRError}
\triangle \tilde{I}'(u,v,\eta) = \triangle \tilde{I}(u,v,\eta) + 
 \sum_{n=0}^{\infty} \left\{P_{n}(\epsilon) \, \epsilon\, FT(\triangle f^{n})\right\}.
\end{equation}
For many foreground sources and subtraction algorithms the first few terms of the Taylor expansion are expected to dominate.

The spherical symmetry discussed here is largely independent of the source subtraction method, and can be used in conjunction with other methods to further reduce the effect of residual foreground contamination.
Because the proposed EOR observations are many orders of magnitude deeper than existing measurements, the brightness and spectral characteristics of many of the foreground sources are poorly constrained. This makes estimating the residual contamination left by particular subtraction algorithms very difficult.  Measuring the foregrounds and studying the residual subtraction errors is the subject of several ongoing research programs and proposed observing campaigns around the world.  

Our current understanding of the residual foregrounds can be summarized as follows.  The smooth spectrum emission from the faint galaxy foreground can be removed using the spectral information, either by using the correlation of visibilities in widely separated frequency channels, or traditional spectral fitting techniques (Zaldarriaga et al. 2004; Briggs et al.\ private communication 2004).  Both of these techniques will clean the faint point source foreground to a level less than the expected EOR signal, and the spherical symmetry can be used to reduce it further.  The smooth spectrum contribution from the Milky Way limits the sensitivity of the observations by being the dominant source of the thermal noise (see Section \ref{SimpleCaseSection}), however it is expected to be quite smooth and easily removed using traditional spectral fitting techniques.  The residual foreground from radio recombination lines is more problematic due to uncertainties in the emission characteristics.  Radio recombination lines in our own galaxy will be measured and subtracted, but care must be taken to accurately determine the relative strength of the different emission lines to allow precision subtraction.  Accurately measuring the Milky Way radio recombination lines is one of the main motivations behind several current observational campaigns.  The radio recombination lines from faint galaxies must be removed to high accuracy because they can mimic the spherical symmetry of the EOR signal, however, they are quite weak.  If radio recombination lines can be subtracted for sources above a few mJy and contribute 0.1\% of the flux for sources weaker than this, their contribution to the total variance will be an order of magnitude fainter than the expected EOR signal.  

The strong symmetry differences between the expected signal and the residual foreground contaminants provide an additional handle for separating the EOR signal from most of the residual foregrounds. In practice, the separation is probably best performed by modeling the residual effects of the foreground sources, and using a maximum likelihood analysis to identify the EOR emission. In this type of analysis the difference in symmetry between the signal and the foregrounds provides one particularly powerful handle, which can be combined other statistical signatures to identify and measure the properties of the EOR. In future work we plan to extensively model the foreground contamination and removal process to fully explore the effect of foreground sources.

\subsection{Non-Gaussian Statistics}
\label{NonGaussianSubSection}

Unlike the density fluctuations of the CMB, the EOR signal is expected to be non-Gaussian in nature. This is because the intensity of the EOR signal is a function of the spin temperature of the gas and the local ionization fraction, in addition to the density fluctuations. While the density fluctuations are largely Gaussian (at least in the linear gravity regime), the spin temperature and ionization fraction may have a very complex distribution \citep{ MMR97,Tozzi}, and the overall statistical distribution is unlikely to have a Gaussian distribution.

The power spectrum is still the Fourier transform of the autocorrelation function and as such is a uniquely important measure of the image statistics, and the development in Section \ref{AnalTechSection} is valid. But unlike with the Gaussian fluctuations of the CMB, the power spectrum of the EOR signal does not fully describe the underlying statistics (see \citet{CoxMiller} and \citet{Vanmarcke} for background on the Weiner-Khinchine theorem and related issues for non-Gaussian random fields).  In particular, the full shape of the statistical distributions within the spherical Fourier annuli---the power spectrum is just the second moment of these distributions---and the global statistical distribution in the image cube may provide important information on the physics of reionization, including the history of the spin temperature and the characteristics of the reionization front.  At this point, there is little theoretical guidance on the shape that the statistical distributions may take, but it is important to include these effects in future calculations as they may provide important information on the process of reionization.

\section{Qualitative Effects of Reionization on the Power Spectrum}
\label{PowerSpectrumSection}

A full prediction of the expected power spectrum and other statistical signatures of the EOR signal will require detailed Monte Carlo simulations.  However, there are a few general features of the EOR power spectrum which are expected. In this section we explore a couple of toy models to help build a conceptual understanding of the power spectrum characteristics and to demonstrate some of the information which can be extracted from the statistics of the EOR signal.   

\begin{figure}
\begin{center}
\plotone{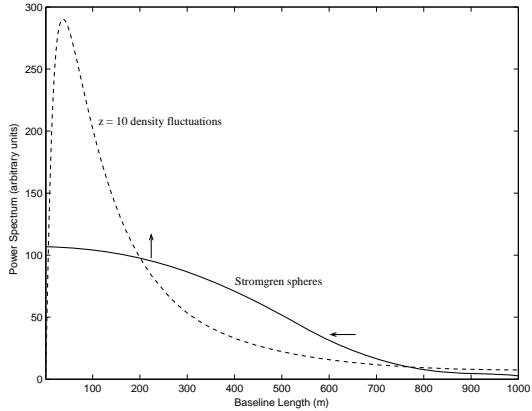}
\caption{Two signatures which are expected to appear in the power spectrum of the EOR signal.  The dashed line is the power spectrum of cosmological density fluctuations (this example is at z = 10).  While the shape of the distribution is robust, particularly at smaller baselines, the amplitude depends on the distribution of spin temperatures and the reionization history.  The solid line is then a cartoon of the power spectrum produced by Stromgren spheres.  The shoulder occurs at the characteristic size of the spheres and moves to the left as they become larger.  At the same time the height of the low $k$ portion will increase, and the ratio between these values can constrain the number of ionizing sources.}
\label{signal}
\end{center}
\end{figure}

The first signature is the power spectrum of matter density fluctuations.  For a simple model with uniform fractional ionization and spin temperature, the EOR signal is simply the primordial power spectrum (see Section \ref{SimpleCaseSection} and \citet{Tozzi}). Generally, the amplitude of this signal depends on the spin temperature, the fractional ionization, and the gravitational growth of density fluctuations.  There are a number of effects which could also distort the observed shape of the density fluctuation power spectrum, but at large spatial scales (small $\vec{k}$) local effects from non-uniform spin temperature and ionization bubbles should have little impact on the shape.  Thus we expect the power spectrum of density fluctuations to be present at some amplitude in all models of the EOR (see Figure \ref{signal}).  One enticing possibility raised by this signature is the opportunity to constrain the cosmological equation of state if the effects of fractional ionization and spin temperature can be separated from the density fluctuation growth term \citep{GnedinEOR1, LoebZalLFT}.

The second signature is due to the Stromgren spheres which are expected to ionize the inter-galactic medium.  These spheres are observed by the absence of neutral hydrogen emission on the ionized interior, and provide a strong power spectrum signal in most models developed to date.  Again, the the shape of small ionized regions may be strongly affected by local phenomena, but are expected to become somewhat spherical on large spatial scales as reionization progresses.  The auto-correlation function of the Stromgren spheres will be strongly centralized, and the Fourier transform will have a $sinc^{2}$ like distribution.  Thus the process of reionization creates an increase in the power spectrum at small $\vec{k}$ with a shoulder at the characteristic length scale of the Stromgren spheres (see Figure \ref{signal} and the simulations by \citet{ZaldariaggaPow1}).  As the spheres enlarge the shoulder will move to smaller $\vec{k}$ values and the amplitude of the power spectrum at small $\vec{k}$ will increase. Note also that the power spectrum will be distinctly different if the universe was ionized by a small number of quasars vs.\ a large number of luminous galaxies.  If a small number of luminous sources dominate reionization, the ratio of the ionization fraction to the characteristic size of the ionized regions will be much lower than if a large number of less luminous sources dominate reionization. By measuring the relative amplitude of the the small $\vec{k}$ component of the power spectrum to the position of the shoulder, we can constrain how many sources participated in reionization of the universe (see Figure \ref{signal}).

These simple examples show the promise of studying the statistical properties of the EOR signal. Very recent results by \citet{RennanNoStep} indicate that reionization was complicated and patchy, and the power on very large length scales is expected to increase significantly. Detailed Monte Carlo predictions of the expected statistical distributions are needed to allow even clearer differentiation between various models of structure formation and reionization.

\section{Simplest Case: Reheated Structures Before Reionization}
\label{SimpleCaseSection}

In this section we use the analysis technique developed in Section \ref{AnalTechSection} to determine the expected signal strength for a typical wide field radio observation.  The model we will consider assumes a preheated spin temperature with no significant ionization and has been extensively discussed by \citet{Tozzi}, \citet{Iliev} and others.  For the instrument we will use the baseline design for the LOFAR observatory and assume a single 100 hour wide field radio observation analyzed using only the power spectrum.\footnote{The full design of the LOFAR observatory (http://www.lofar.org) is currently in a state of flux, and it is unclear whether the observatory being constructed in the Netherlands will be capable of wide field observations or have the full baseline sensitivity.  However, this is a good baseline and future observatories are likely to be significantly more sensitive.}  This case is useful for comparing with pervious work and sets a baseline sensitivity for this analysis technique.  The exact sensitivity of any measurement depends intimately on the characteristics of the instrument. However, an estimate of the EOR sensitivity can be developed using a few approximations. 

\begin{figure}
\begin{center}
\plotone{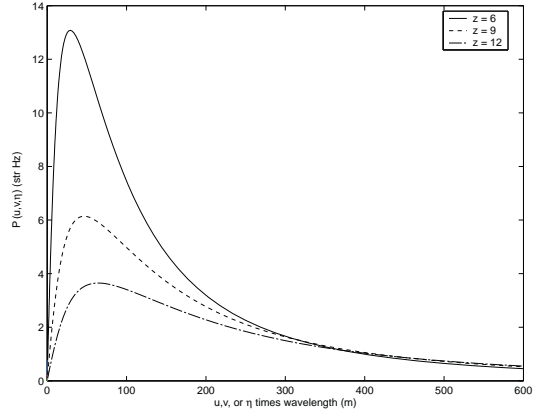}
\caption{The power spectrum of matter density fluctuations in observer coordinates $u, v, \eta$, with no ionization.  The 21 cm emission has been redshifted to 203 MHz, 142 MHz, and 109 MHz for z = 6, 9, 12 respectively.  The horizontal axis has been scaled by the wavelength so that the visibility measurements made by antenna pairs remain constant in position for all three redshifts.  In the current LOFAR design, the visibility measurements from the central core extend to distances of 2 km, though the distribution of these measurements has not been determined.}
\label{PowerSpectrum}
\end{center}
\end{figure}

For observations at 129 MHz (z=10) with a uniformly heated medium ($T_{\rm s} \gg T_{\rm CMB}$), $P_{\rm HI}$ is just the constant $C_{\rm Jy}^{2} \approx 164$  Jy$^{2}$ str$^{-2}$ times the matter density power spectrum (see Appendix \ref{AppendixUnits} for a discussion of units).  The power spectrum of the matter density fluctuations is shown in Figure \ref{PowerSpectrum}, indicating the strong signature of the density fluctuations which will be imprinted on the EOR signal. The function $|W(u,v,\eta)|^{2}$ from Equation \ref{pkconclusion} is very sharply peaked, so we can make the approximation that the power spectrum is a constant on the length scale of $|W(u,v,\eta)|^{2}$ and pull the matter density power spectrum out of the integral. (Since in practice we are fully sampling the instrumental bandwidth, we are really performing a uniform average over $\eta$ in this argument and an equivalent answer can be obtained using a variation of Parseval's relation for $\eta$.)  If the response of the antennas can be described as constant to $20^{\circ}$ with 4 MHz bandwidth, the integral of $|W(u,v,\eta)|^{2}$ is $3.8 \times 10^{5}$ str Hz by Parseval's relation. The baseline LOFAR design is centrally condensed and does not evenly sample the $u, v, \eta$ space, but the magnitude of the condensation has yet to be determined.  If we assume that the sampling varies as approximately as $r^{-2}$ where $r=\sqrt{u^{2}+v^{2}+\eta^{2}}$, the average neutral hydrogen power spectrum is 0.74 str Hz.  Combining all of these factors we obtain an average $\left\langle \lvert \triangle \tilde{I}(u_{1},v_{1},\eta_{1}) \rvert^{2} \right\rangle$ signal of $4.5 \times 10^{7}$ Jy$^{2}$ Hz$^{2}$. 

The next step in calculating the sensitivity is to determine the thermal noise inherent in the measurements.  The total thermal noise also depends on the distribution of antennas on the ground, but for a dense array may be approximated by the following argument.  For a single dual polarization visibility measurement from a pair of high frequency antenna elements (each with 16 dipoles) the thermal noise is given by
\begin{equation}
\label{ }
\triangle V_{\rm rms} = \frac{\sqrt{2} k_{B}T_{\rm sys}}{4\lambda^{2}\sqrt{df\, \tau}},
\end{equation}
where $df$ is the width of the individual frequency channels, and $\tau$ is the duration of the observation.  To calculate the noise on $\triangle \tilde{I}(u,v,\eta)$, we need to Fourier transform $\triangle V(u,v,f)$.  Since the noise in $\triangle \tilde{I}(u,v,\eta)$ will be the same at all $\eta$, we can calculate the noise at $\eta = 0$ and generalize to the other values.  $\triangle \tilde{I}(u,v,0)$ becomes
\begin{subequations}
\label{ }
\begin{align}
\label{}
   \triangle \tilde{I}(u,v,0)& = \int V(u,v,f)\, df,   \\
      &=\sum^{B/df} V(u,v,f)\, df,
\end{align}
\end{subequations}
where $B$ is the full bandwidth of the measurement.  The rms of the sum is $\sqrt{N} = \sqrt{B/df}$ times the rms of $V$ times $df$, and is given by
\begin{equation}
\label{ }
\triangle \tilde{I}_{\rm rms} = \frac{\sqrt{2} k_{B}T_{sys}\sqrt{B}}{4 \lambda^{2}\sqrt{\tau}}.
\end{equation}
Using the values of $T_{\rm sys}$ = 695 K, $\lambda$ = 2.32 m, $\tau$ = 100 h, and $B$ = 4 MHz gives $\triangle \tilde{I}_{\rm rms} = 2.1 \times 10^{5}$ Jy Hz. 

The background noise $ \langle \triangle \tilde{I}(u,v,\eta)  \rangle$ is Gaussian distributed with rms $\sigma$.  The expectation of the absolute value $\langle |\triangle \tilde{I}(u,v,\eta)|\rangle$ is then Rayleigh distributed and the expectation of the absolute value squared $\langle|\triangle \tilde{I}(u,v,\eta)|^{2}\rangle$ is exponentially distributed with an rms of $2\sigma^{2}$.  This gives an rms for a single $|\triangle I(u,v,\eta)|^{2}$ measurement of $8.8 \times 10^{10}$ Jy$^{2}$ Hz$^{2}$.  If we average over a large number of visibilities, the mean  $\overline{\langle|\triangle \tilde{I}(u,v,\eta)|^{2}\rangle}$ will become Gaussian distributed with width rms/$\sqrt{N_{\rm vis}}$, where $N_{\rm vis}$ is the number of visibility measurements in the sum. 

For our example we have 1.2 billion visibility measurements from the wide field correlator, giving an rms of $2.5 \times 10^{6}$ Jy$^{2}$ Hz$^{2}$ for the thermal noise. Comparing to the expected signal strength of $4.5 \times 10^{7}$ Jy$^{2}$ Hz$^{2}$ results in a signal to noise ratio of $\sim 15$ for a single 100 h 4 MHz observation in the absence of significant foreground contamination, or equivalently a $\sim 3\sigma$ signal in 5 separate annular bins.  This suggests that precision measurements of the power spectrum can be achieved using wide field radio observations and reasonable integrations.


\section{Conclusion}

In this paper we discussed the power of wide field radio observations and introduced an analysis technique which we expect will efficiently extract the statistical properties of the EOR signal and separate them from the residual foreground contamination.  The next step in developing the experimental procedure is to use simulations to explore a number of subtle experimental effects and their impact on EOR observations.  Using end-to-end simulations which include realistic foreground models and instrumental effects, we intend to explore four major topics:
\begin{itemize}
  \item Foreground contamination and our ability to measure the EOR power spectrum using various foreground subtraction methods and symmetry techniques.
  \item Ionospheric effects and calibration errors, and the impact on deep EOR integrations.
  \item Array optimization to enhance the science return of EOR measurements in future observatories.
  \item Statistical properties of theroetical models of reionization, and the ability of wide field radio observations to differentiate between EOR models.
\end{itemize}
It is clear that wide field radio observations can make an important contribution to the study of the early universe, and the above steps will help formulate a robust strategy for measuring and interpreting the EOR signal.

\section{Acknowledgments}
There are many people who have provided important contributions to this paper.  Ed Bertschinger provided valuable feedback on early drafts of this paper, taught MFM the basics of cosmological statistics, and pointed out the importance of non-Gaussian statistics.  Numerous other people have provided valuable feedback on various drafts and helped us develop our ideas through extensive discussions, including Colin Lonsdale, Ger de Bruyn, Frank Briggs, Rennan Barkana, Al Levine, and Judd Bowman.  We also thank our low frequency radio team members at MIT, Haystack observatory, ASTRON, and NRL. This work has been supported by NSF grant AST0121164.

\setcounter{equation}{0}  
\setcounter{section}{0}
\renewcommand{\thesection}{\Alph{section}}
\renewcommand{\theequation}{\Alph{section}-\arabic{equation}}

\section{Derivation of Brightness Temperature}
\label{AppendixBD}

The optical depth $\tau_{\nu}$ is related to the absorption coefficient $\alpha_{\nu}$ by (see, for example, \citet{RL} Equation 1.26)
\begin{equation}
\label{ODdef}
\tau_{\nu}(s) =\int_{s_{0}}^{s}\alpha_{\nu}(s')ds'.
\end{equation}
Under conditions of local thermodynamic equilibrium applicable for a thermalized neutral hydrogen gas the absorption coefficient is given by
\begin{equation}
\label{ }
\alpha_{\nu}=\frac{h\nu}{4\pi}n_{1}B_{12}\left(1-e^{-h\nu/kT_{s}}\right)\phi(\nu)
\end{equation}
from \citet{RL} Equation 1.80, where $\nu$ is the frequency, $n_{1}$ the number density of lower quantum states, $B_{12}$ the Einstein absorption coefficient, $T_{s}$ is the temperature of the spin states of the neutral hydrogen and $\phi(\nu)$ is the line profile. The relation between the absorption coefficient $B_{12}$ and the transition probability for spontaneous emission $A_{21}$ with the number of states in the upper and lower energy levels given by $g_{2}$ and $g_{1}$ respectively is
\begin{equation}
\label{ }
A_{21}= \frac{2h\nu^{3}g_{1}}{c^{2}g_{2}}B_{12},
\end{equation}
which we can then use to obtain
\begin{equation}
\label{ }
\alpha_{\nu}=\frac{c^{2}}{8\pi\nu^{2}}\frac{g_{2}}{g_{1}}n_{1}A_{21}\left(1-e^{-h\nu/kT_{s}}\right)\phi(\nu).
\end{equation}

For the 21 cm line of hydrogen, the factor $h\nu/k$ is 0.068 K.  This is a much lower temperature than the CMB radiation, and for almost all cases the spin temperature $T_{s}$ will be warmer than this even when the 21 cm line is seen in absorption.  Making the approximation $h\nu/k\ll T_{s}$ we find
\begin{equation}
\label{ }
\alpha_{\nu}=\frac{c^{2}}{8\pi\nu^{2}}\frac{g_{2}}{g_{1}}n_{1}A_{21}\frac{h\nu}{kT_{s}}\phi(\nu).
\end{equation}
Using the same approximation we can also show that the hydrogen will be equally distributed between all of the $s_{1}$ quantum states.  For the 21 cm transition $g_{2}=3$ and $g_{1}=1$.  This then implies that the $n_{1}=n_{\rm HI}/4$, where $n_{\rm HI}$ is the number density of neutral hydrogen at the given redshift and implicity includes the density fluctuations (the units have not been defined yet).  Inserting these values we obtain
\begin{equation}
\label{ }
\alpha_{\nu}=\frac{3c^{2}}{32\pi\nu^{2}}n_{\rm HI}A_{21}\frac{h\nu}{kT_{s}}\phi(\nu).
\end{equation}

Now to calculate the optical depth we need to perform the integration in Equation \ref{ODdef}. Because the 21 cm line is extremely narrow all of the optical depth is constrained to a specific redshift $z$.  Effectively $\phi(\nu)$ becomes a $\delta$-function over the line of sight variable $s'$.  We can thus rewrite the equation using $\phi(\nu)=\delta(\nu_{21}-(1+z')\nu)=\frac{(1+z')}{\nu_{21}}\delta(z-z')$ and perform the integral to obtain
\begin{eqnarray}
\tau_{\nu} &= & \int \frac{3c^{2}}{32\pi\nu_{21}^{2}}n_{\rm HI}A_{21}\frac{h\nu_{21}}{kT_{s}}\frac{(1+z')}{\nu_{21}}\delta(z-z') \frac{ds}{dz'}dz'\\
\tau_{\nu} &= &\frac{3c^{2}}{32\pi\nu_{21}^{2}}n_{\rm HI}A_{21}\frac{h\nu_{21}}{kT_{s}}\frac{(1+z)}{\nu_{21}}\frac{ds}{dz},
\label{ODfinal}
\end{eqnarray}
where we have evaluated the frequency terms from the transition strength at the 21 cm line and left the $ds/dz$ conversion from length scale to $z$ until we explicitly decide on units in Equation \ref{unitstep} below.

We can now use the optical depth to obtain the specific intensity at the source of the 21 cm emission.  From \citet{RL} Equation 1.30 the specific intensity after the emission region ($I_{v}$) is given in terms of the background brightness $I_{\rm CMB}$, which is dominated by the cosmic microwave background, and the emission brightness of the neutral hydrogen $I_{\rm HI}$ via
\begin{equation}
\label{ }
I_{v}=I_{\rm CMB}e^{-\tau_{\nu}} + I_{\rm HI}\left(1-e^{-\tau_{\nu}}\right).
\end{equation}
The difference between the background and the neutral hydrogen emission is then given by
\begin{equation}
\label{Bfinal}
\bigtriangleup I_{v}=\left(I_{\rm HI}-I_{\rm CMB}\right)\left(1-e^{-\tau_{\nu}}\right) \approx \left(I_{\rm HI}-I_{\rm CMB}\right)\tau_{\nu},
\end{equation}
where the last relation follows from the small optical depths associated with the 21 cm line.

The final step in the process is to propagate the radiation from the emission region to the Earth.  The number density $n_{\rm HI}$ is related to the mean hydrogen density $\langle n_{\rm H}\rangle$ times the neutral hydrogen density fluctuations
\begin{equation}
\label{ }
n_{\rm HI}=\left\langle n_{\rm H}\right \rangle \frac{\rho_{\rm HI}}{\langle\rho_{\rm H}\rangle}.
\end{equation}
We now have to decide on the units we will use for $n_{\rm HI}$, $ds/dz$ and the relativistic propagation relationship.  Here we choose the units of the emission region to be proper Mpc, though all unit choices will give the same answer.  Since the average neutral hydrogen density is locked to the Hubble flow, the mean density evolves as $(1+z)^{3}$.  The mean hydrogen density can then be defined in terms of the present value of the cosmological baryon density $\Omega_{\rm B}$ to obtain
\begin{equation}
\label{unitstep}
\langle n_{\rm H} \rangle = \frac{\rho_{B}}{m_{\rm H}}(1-Y) (1+z)^{3}= \left(\Omega_{\rm B}\, h^{2}\right)(1-Y)\frac{3(H_{0}/h)^{2}}{8\pi\,G\,m_{\rm H}}(1+z)^{3},
\end{equation}
where $Y$ is the helium fraction, $G$ is the gravitational constant, $\rho_{B}$ is the physical baryon density, $H_{0}/h$ is the Hubble constant, and $m_{\rm H}$ is the mass of a hydrogen atom.  The proper length per redshift interval $ds/dz$ is given by
\begin{equation}
\label{ }
\frac{ds}{dz}= \frac{c}{H_{0}\,(1+z)E(z)},
\end{equation}
where $E(z) \equiv \sqrt{\Omega_{\rm M}(1+z)^{3} + \Omega_{\rm k}(1+z)^{2} + \Omega_{\Lambda}}$.  In proper coordinates the brightness conservation law is given by (from \citet{PeacockBook} Equation 3.88)
\begin{equation}
\label{Bevolve}
I_{\nu}(0)=\frac{I_{v}(z)}{(1+z)^{3}}.
\end{equation}
Putting together Equations \ref{ODfinal} and \ref{Bfinal} through \ref{Bevolve} and simplifying one then obtains
\begin{multline}
\label{HIBrightness}
\bigtriangleup I_{\nu}= \frac{3c^{3}}{32\pi \nu_{21}^{3}} \left[\left(\Omega_{\rm B}\, h^{2}\right)(1-Y)\frac{3(H_{0}/h)^{2}}{8\pi\,G\,m_{\rm H}}\right] \times \\ \frac{A_{21}}{H_{0}\,E(z)} \frac{h\nu_{21}}{k}\frac{\left(I_{\rm HI}-I_{\rm CMB} \right)}{T_{s}}\frac{\rho_{\rm HI}}{\langle\rho_{\rm H}\rangle}.
\end{multline}
The specific intensities can be replaced by brightness temperatures to obtain
\begin{multline}
\label{ }
\bigtriangleup T_{\nu}= \frac{3c^{3}}{32\pi \nu_{21}^{3}} \left[\left(\Omega_{\rm B}\, h^{2}\right)(1-Y)\frac{3(H_{0}/h)^{2}}{8\pi\,G\,m_{\rm H}}\right] \times \\ \frac{A_{21}}{H_{0}\,E(z)}  \frac{h\nu_{21}}{k}\frac{\left(T_{s}-T_{\rm CMB} \right)}{T_{s}}(1+z)^{2}\frac{\rho_{\rm HI}}{\langle\rho_{\rm H}\rangle},
\end{multline}
where $T_{s}$ is the spin temperature and the $(1+z)^{2}$ factor relates the temperature on the left hand side which is measured at the antenna and the temperatures on the right hand side which are in the frame of the neutral hydrogen. 

Substituting standard values for the constants ($\Omega_{\rm B}h^{2}$ = 0.02, $Y$ = 0.24) these simplify to
\begin{equation}
\label{ }
\bigtriangleup I_{\nu}= (2.9\ {\rm mK})\,h^{-1}\frac{\left(I_{\rm HI}-I_{\rm CMB} \right)}{E(z)\, T_{s}}\frac{\rho_{\rm HI}}{\langle\rho_{\rm H}\rangle},
\end{equation}
\begin{equation}
\label{ }
\bigtriangleup T_{\nu}= (2.9\ {\rm mK})\,h^{-1}\frac{(1+z)^{2}} {E(z)}\frac{\left(T_{\rm s}-T_{\rm CMB} \right)}{T_{s}}\frac{\rho_{\rm HI}}{\langle\rho_{\rm H}\rangle}.
\end{equation}

\setcounter{equation}{0}  

\section{Units}
\label{AppendixUnits}

The measured surface brightness in sky coordinates $\triangle I(\theta_{x},\theta_{y},f)$ is in Jy str$^{-1}$.  The window function in sky coordinates $W(\theta_{x},\theta_{y},f)$ is unitless. The Fourier transformed surface brightness $\triangle \tilde{I}(u,v,\eta)$ is in Jy Hz, and the window function $W(u,v,\eta)$ is in str Hz.  For equations such as number \ref{trueIconv} the convolution over $du\ dv\ d\eta$ removes the extra factor of str Hz (see Footnote \ref{FTnote}). The measured visibility $V(u,v,f)$ is measured in Jy.  

The quantities $C_{\rm Jy}$ and $C_{\rm K}$ are measured in Jy str$^{-1}$  and K respectively. The neutral hydrogen density $\frac{\rho_{\rm HI}-\langle\rho_{\rm H}\rangle}{\langle\rho_{\rm H}\rangle}$ is unitless in $\theta_{x}, \theta_{y}, \triangle f$ coordinates, and has units of str Hz in Fourier coordinates.  The delta function $\delta(\vec{n}-\vec{n}''')$ has units of str$^{-1}$ Hz$^{-1}$, leading to the power spectrum $P_{\rm HI}(\vec{n})$ having units of Jy$^{2}$ Hz str$^{-1}$ or K$^{2}$ str Hz.  If the spin temperature is a constant, $C$ can be removed and the power spectrum of the neutral hydrogen density fluctuations is str Hz as one would expect.

 \end{document}